\documentclass[prl,twocolumn,superscriptaddress]{revtex4}

\usepackage{bm}
\usepackage{graphics}
\usepackage{amsmath,epsfig}
\usepackage{amssymb}
\begin{document}

\title{Experimental Construction of Optical
Multi-qubit Cluster States From Bell States }

\author{An-Ning Zhang}
\affiliation{Department of Modern Physics and Hefei National
Laboratory for Physical Sciences at Microscale, University of
Science and Technology of China, Hefei, Anhui 230026, People's
Republic of China}
\author{Chao-Yang Lu}
\affiliation{Department of Modern Physics and Hefei National
Laboratory for Physical Sciences at Microscale, University of
Science and Technology of China, Hefei, Anhui 230026, People's
Republic of China}
\author{Xiao-Qi Zhou}
\affiliation{Department of Modern Physics and Hefei National
Laboratory for Physical Sciences at Microscale, University of
Science and Technology of China, Hefei, Anhui 230026, People's
Republic of China}
\author{Yu-Ao Chen}
\affiliation{Physikaliches Institut, Universit\"{a}t Heidelberg,
Philisophenweg 12, 69120 Heidelberg, Germany}
\author{ Zhi Zhao}
\affiliation{Department of Modern Physics and Hefei National
Laboratory for Physical Sciences at Microscale, University of
Science and Technology of China, Hefei, Anhui 230026, People's
Republic of China} \affiliation{Physikaliches Institut,
Universit\"{a}t Heidelberg, Philisophenweg 12, 69120 Heidelberg,
Germany}
\author{Tao Yang}
\affiliation{Department of Modern Physics and Hefei National
Laboratory for Physical Sciences at Microscale, University of
Science and Technology of China, Hefei, Anhui 230026, People's
Republic of China}
\author{Jian-Wei Pan}
\affiliation{Department of Modern Physics and Hefei National
Laboratory for Physical Sciences at Microscale, University of
Science and Technology of China, Hefei, Anhui 230026, People's
Republic of China} \affiliation{Physikaliches Institut,
Universit\"{a}t Heidelberg, Philisophenweg 12, 69120 Heidelberg,
Germany}

\date{\today}

\begin{abstract}
Cluster states serve as the central physical resource for the
measurement-based quantum computation. We here present a simple
experimental demonstration of the scalable
cluster-state-construction scheme proposed by Browne and Rudolph.
In our experiment, three-photon cluster states are created from
two Bell states using linear optical devices. By observing a
violation of three-particle Mermin inequality of $|\langle
\textit{A}\rangle| = 3.10\pm0.03 $, we also for the first time
report a genuine three-photon entanglement. In addition, the
entanglement properties of the cluster states are examined under
$\sigma_z$ and $\sigma_x$ measurements on a qubit.

\end{abstract}

\pacs{Valid PACS appear here}

\maketitle

There has been considerable interest in optical approaches to
quantum computation due to photon's intrinsic robustness against
decoherence and the relatively ease of manipulation with high
precision. Remarkably, by exploiting the nonlinearity induced by
measurement, Knill, Laflamme, and Milburn showed that efficient
quantum computation is possible with linear optics \cite{klm01}. A
number of simplifications and modifications \cite{yr03, nielsen04}
of this scheme, as well as experimental demonstrations of the most
elementary components \cite{opw03, sjp03, gpw04, zzc04} have been
reported.

Surprisingly, Raussendorf and Briegel proposed a conceptually new
quantum computation model \cite{rb01}. They have shown that
universal quantum computation can be done by one-qubit
measurements on a specific entangled state, the cluster state
\cite{br01}. With the cluster states prepared, information is then
written onto, processed, and read out from the cluster by
one-particle measurement only. After a sequence of one-qubit
measurement which forms the computational program, the
entanglement in a cluster state is destroyed. Therefore this
scheme was called as ``one-way quantum computer'' or
``measurement-based quantum computer''. Underlying this novel
computation model is the cluster states, serving as the entire
physical resources. Many efforts have been devoted to constructing
the cluster states. Proposals and experiments using neutral atoms
trapped in the periodic potential of an optical lattice with
controlled collisions between neighboring atoms have been reported
\cite{jbc99}. On the optical approach, Nielsen \cite{nielsen04}
showed that optical cluster states can be efficiently created
using non-deterministic gates from the KLM scheme. Recently, a
much more simple and powerful linear optical quantum computation
scheme was proposed by Browne and Rudolph \cite{br04}. They showed
how cluster states may be efficiently generated from pairs of
maximally entangled photons in a scalable way using some technique
called qubits ``fusion''. It is significantly less demanding not
only in resource requirement, but also in complexity of
experimental implementation.

In this letter, we report the first experimental demonstration of
constructing linear multi-qubit cluster states from pairs of Bell
states. As the most fundamental step in Browne and Rudolph's
scheme, we produced a three-photon cluster states by a Type-I
``fusion'' of two pairs of maximally polarized entangled photons.
We then provide sufficient experimental evidence confirming that
the cluster states we obtained, unitarily equivalent to the
three-photon Greenberger-Horne-Zeilinger (GHZ) states
\cite{ghz89}, are genuine three-particle entanglement, thus
excluding any possibility of hybrid models \cite{su01}. We also
examined the entanglement properties of the remaining two photons
under a measurement on a qubit in different basis $\{\sigma_z,
\sigma_x\}$.

\begin{figure}[b]
  \begin{center}
  \includegraphics[width=2.8in]{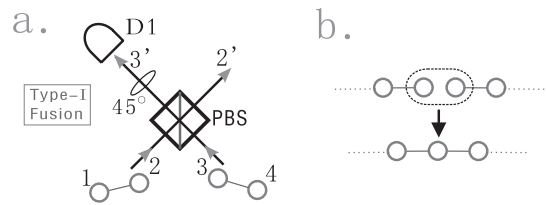}
  \end{center}
  \caption{
  (a). non-deterministic qubit ``fusion'' operation. D1 stand for a polarization discriminating photon detector.
  Two photons of different spatial modes are mixed in a PBS,
  the output in mode $3'$ is accepted only when D1 receives exactly one photon.
  (b). A success Type-I ``fusion'' combines two linear clusters of length n and m into a new one of length (n+m-1).}
  \label{fig:fusion}
\end{figure}

Let us first review Browne and Rudolph's efficient linear optical
quantum computation scheme. The primary resource used is two
photon Bell states which are relatively easier to obtain,
probabilistically from single photons for example. Given a supply
of Bell states, arbitrarily long linear cluster states can be
generated efficiently using an operation called Type-I qubit
``fusion'' (see Fig.1(a)). This operation, the same as parity
check \cite{psb01,pittman01}, is implemented by mixing two photons
in a polarizing beam splitter (PBS) and accepting the output in
mode $3'$ only for those cases in which polarization-sensitive
detector D1 receives one and only one photon. Here is the most
simple and fundamental case: from two pairs of Bell states to a
three-qubit cluster state. Encoded in polarization, a Bell state,
also equivalent to a two-qubit cluster states under a local
unitary transformation can be written as:
$$|\phi\rangle_{ij}=|H\rangle_i|H\rangle_j+|V\rangle_i|V\rangle_j
=_{\textit{l.u.}}|H\rangle_i|+\rangle_j+|V\rangle_i|-\rangle_j$$

Here H and V denote horizontal and vertical polarizations,
$|\pm\rangle=\frac{1}{\sqrt{2}}|H\rangle\pm|V\rangle$; $i$ and $j$
index the photon's spatial modes. Given two pairs of Bell states,
$|\phi\rangle_{12}$ and $|\phi\rangle_{34}$, we then superpose
photon 2 and 3 in a PBS. Since the PBS transmits horizontal and
reflects vertical polarization, detecting one and only one photon
in D1 makes sure that both photon 2 and photon 3 are horizontally
polarized or vertically polarized. By a further measurement
performing on output $2'$ in the $+/-$ basis, photon 1, 3, 4 will
be in a three-photon cluster state:
$$|\phi^{\pm}\rangle_{134} = |+\rangle_1|H\rangle_3|+\rangle_4 \pm
|-\rangle_1|V\rangle_3|-\rangle_4 $$ depending on the measurement
result of detector D1. It is equivalent to a three-qubit GHZ state
\cite{ghz89} under local unitary transformation.

As has been discussed in detail by Browne and Rudolph in Ref.
\cite{br04}, the creation of three-photon cluster states is the
most fundamental step towards the goal of constructing a square
lattice cluster state that would allow a simulation of arbitrary
quantum network directly by single-qubit measurement alone
\cite{rb01}. With a success probability of $50\%$, Type-I
``fusion'' combined two linear cluster state of lengths n and m
into a new one of length (n+m-1). Any linear cluster states with
desired length can be efficiently created with this method given
necessary resource of Bell states. Further, we can also generate
arbitrary two-dimensional cluster from those obtained linear
cluster states by some similar methods \cite{detail}.

\begin{figure}[t]
  \begin{center}
  \includegraphics[width=3.0in]{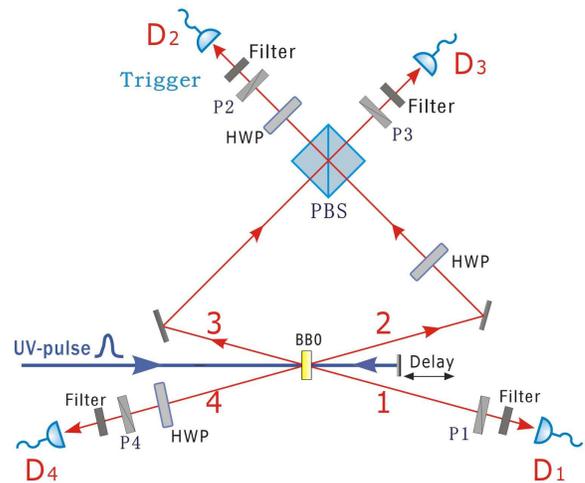}
  \end{center}
  \caption{Experimental setup of generating three-photon cluster state from two pairs of maximally entangled photons produced by Type-II spontaneous parametric down-conversion. The half wave plate (HWP) used in path 2 and 4 are used to locally transform the photon from $H/V$ basis to $+/-$ basis and the four polarizers P1, P2, P3, P4 are used for necessary polarization analysis. In the experiment, we managed to obtain an average twofold coincidence of $2.2\times10^4 s^{-1}$. }
  \label{fig:cluster}
\end{figure}

We note that, compared to the previous schemes, it not only
reduces the resources required but also moves away the difficulty
of interferometric phase stability. And there are also further
advantages reported in Ref. \cite{br04}. Obviously, given perfect
photon pairs and number-discriminating photon detectors, the
scheme described above can be realized optimally without
postselection. However, we note that, although these techniques
are not available yet, it is still sufficient to perform an
experimental demonstration based on postseletion.

A schematic drawing of our experimental setup is shown in Fig.2.
An ultraviolet pulsed laser from a mode-locked Ti:sapphire laser
(center wavelength 394nm, pulse duration 200fs, repetition rate
76MHz) passed through  $\beta$-barium borate (BBO) crystal twice
to generate two maximally entangled photon pairs in mode 1-2 and
mode 3-4. After proper birefringence compensation and local
unitary transformation with half wave plate (HWP) and nonlinear
crystals, two pairs of two-qubit cluster states are produced as
the primary source.

We then superpose the photon 2 and photon 3 at the PBS. Their path
lengths are adjusted such that they arrive simultaneously. To
achieve good spatial and temporal overlap, the outputs are
spectrally filtered $(\triangle\lambda=2.8nm)$ and monitored by
fiber-coupled single-photon detectors. The filtering process
stretches the coherence time to about $740 fs$, substantially
larger than the pump pulse duration \cite{zzw95}. There processes
effectively erase any possibility to distinguish the two photons
and subsequently lead to interference.

\begin{figure}[t]
  \begin{center}
  \includegraphics[width=3.1in]{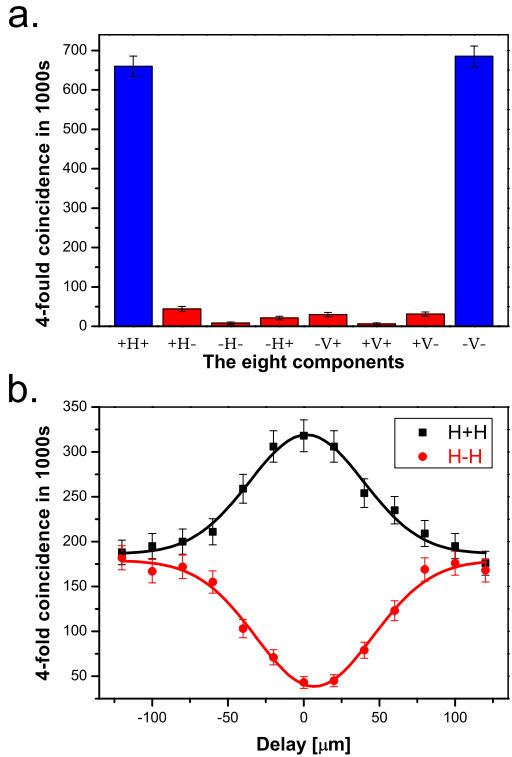}
  \end{center}
  \caption{(a) Experimental data under $8$ different polarizing settings. Two desired terms $+H+$ and $-V-$ are prominent while other six are strongly depressed to be about $3\%$ of any desired ones on average. (b) Experimental data in the ``diagonal'' basis showing the two components are in a coherent superposition. Maximum interference occurs at zero delay between the two incoming photons.}\label{fig:setup}
\end{figure}

To experimentally verify the three-photon cluster state, we first
show that, upon a trigger of D2, the three-fold coincidence only
includes $+H+$ and $-V-$ components, but no others. This is done
by comparing the counts of all $8$ possible polarization
combination $+H+,\cdots, -V-$. The experimental results in the
$(+/-,H/V,+/-)$ basis (see Fig.3(a)) show that the signal-to-noise
ratio defined as the ratio of any of the desired threefold events
($+H+$ and $-V-$) to any of the six other undesired ones is about
$29:1$ on average. Second, we further perform a polarization
measurement in the ``diagonal'' basis $(H/V,+/-,H/V)$ to
demonstrate that the two terms $+H+$ and $-V-$ are indeed in a
coherent superposition. Transform $|\phi^+\rangle_{123}$ in the
``diagonal'' basis $(H/V,+/-,H/V)$, we note that only components
$(H+H, H-V, V+V, V-H)$ occur, other combinations $(H-H, H-V, V+H,
V-V)$ do not occur. As a test for coherence, we compare the H+H
and H-H count rates as a function of the pump delay mirror
position. It shows in Fig.3(b) that, at zero delay, the unwanted
component is suppressed with a visibility of $0.78\pm0.03$, which
is sufficient to violate the Bell-type inequality imposed by local
realism \cite{zk97}.

However, as in the previous experiments of Bouwmeester \textit{et
al.} \cite{dik99} and Pan \textit{et al.} \cite{pan00}, the data
presented above are still not sufficient to confirm the genuine
entanglement of all three particles \cite{su01}. This has been
shown by M. Seevinck and J. Uffink that it can be explained by a
hybrid model in which only less than three particles is entangled.
Aim to exclude such a hybrid model and produce the three-photon
GHZ state in the form $|HHH\rangle+|VVV\rangle$, we first did a
local transformation of the cluster state and
 performed four series of measurements in the
  $\sigma_x\sigma_x\sigma_x, \sigma_x\sigma_y\sigma_y, \sigma_y\sigma_y\sigma_x$
and $\sigma_y\sigma_x\sigma_y$ direction. We then test the
three-particle Bell inequality of the form derived by Mermin
\cite{mermin90} and the result shows:
$$|\langle \textit{A}\rangle| = 3.10\pm0.03 $$
where
$$\textit{A} =
\sigma_x\sigma_y\sigma_y+\sigma_y\sigma_x\sigma_y+\sigma_y\sigma_y\sigma_x-\sigma_x\sigma_x\sigma_x$$

It clearly shows a violation of the inequality: $|\langle
\textit{A}\rangle| \leqslant 2$ imposed by local realism by 34
standard deviations. As has been discussed in Ref. \cite{su01},
confirmation of genuine three-particle entanglement requires a
violation of inequality: $|\langle \textit{A}\rangle| \leqslant
2\sqrt2$. The experimental result also well exceeds the bound to
confirm genuine three-photon entanglement, with a violation of
this inequality by 11 standard deviations, hence leads to
verification of genuine three-photon entanglement. Our
three-photon entanglement source thus distinguishes itself from
all previous ones by its high purity, which would make it possible
to perform a lot of quantum information processing tasks, such as
quantum secreting sharing and the third-man cryptography
\cite{hbb99, thirdman}.

\begin{figure}[b]
  \begin{center}
  \includegraphics[width=3.0in]{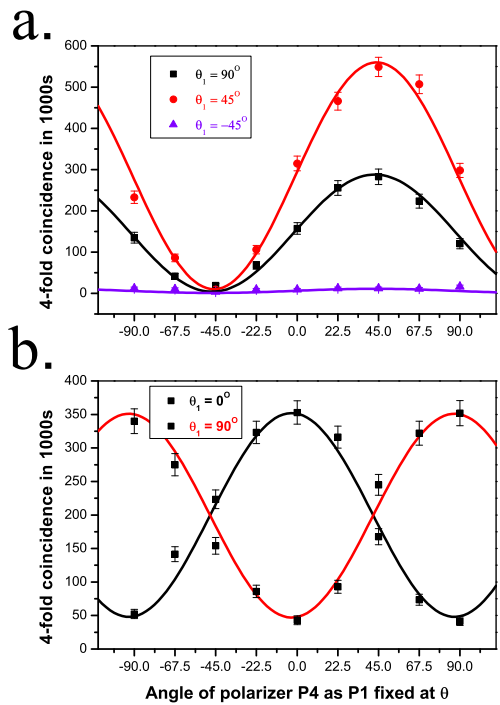}
  \end{center}
  \caption{Experimental results showing polarization correlation between photon 1 and 4, under a $\sigma_z$ and $\sigma_x$ measurement of photon 3.
  (a). Data obtained under a $\sigma_z$ measurement. The coincident counts when P1 was set at $-45^\circ$ was so strongly suppressed that they can hardly concerned; while counts when P1 was set at $45^\circ$ are the most prominent, twice as when P1 was set at $90^\circ$. The experimental results clearly agree that what we obtained is $|+\rangle_1\otimes|+\rangle_4$.
  (b). Data obtained under a $\sigma_x$ measurement. The two sinusoidal curves with a visibility of $0.79\pm0.03$ demonstrate that photon 1 and 4 are in an entangled state as $|+\rangle_1|+\rangle_4+|-\rangle_1|-\rangle_4$.}\label{fig:measurement}
\end{figure}
A quite interesting entanglement property of a linear cluster
state is that, measurements in $\sigma_z$ and $\sigma_x$ basis on
a qubit of a cluster state have totally different effects on the
remaining qubits. This has been shown in Ref. \cite{br04} that, a
$\sigma_z$ eigenbasis measurement removes the qubit from the
cluster sand breaks all bond between that qubit and the rest of
the cluster; while a $\sigma_x$ measurement on a linear cluster
removes the measured qubit and it combines the adjacent qubits
into a redundantly encoded qubit. It is quite critical for us to
understand the cluster-state-construction scheme \cite{br04} and
the cluster model of quantum computation \cite{rb01}. We then
examined the entanglement properties of the two remaining photons
under a $\sigma_z$ measurement and a $\sigma_x$ measurement on the
``mid'' qubit upon trigger of detecting a $|H\rangle$ photon and a
$|+\rangle$ photon by D3 respectively. We analyzed the
polarization correlations between photon 1 and 4 by keeping
polarizer 1 fixed and varying the angle of polarizer 4. The
experimental results are shown in Fig.4(a) and Fig.4(b)
corresponding to measurement in the $\sigma_z$ and $\sigma_x$
basis respectively. As Fig.4 shows, the experimental data are in
good agreement with theoretical prediction.

In summary, we have demonstrated the process of constructing
linear three-photon cluster state from two Bell states. In
principle these method can be extended to any desired number of
particles given enough Bell states, which holds the promise of
constructing an optical one-way quantum computer efficiently. Our
experiment can also be considered as a demonstration of producing
a genuine three-photon GHZ state \cite{pdg01} in an event-ready
way, which in principle does not need postselection given perfect
photon pairs and perfect detectors. The genuineness of the
three-photon entangled state was confirmed by violating the
inequality: $|\langle \textit{A}\rangle| \leqslant 2\sqrt2$ by 11
standard deviations. After verification of the obtained
three-photon cluster state, we also demonstrate that a $\sigma_z$
measurement on a qubit of the obtained three-photon cluster state
breaks the bond between the rest photons; while a $\sigma_x$
measurement does not, but instead combines them into a redundantly
encoded qubit. However, in this experiment, only partial features
of cluster states and the cluster-state-construction scheme were
demonstrated. Possible future work could include production and
characterization of  a four-photon cluster state, which is local
unitary inequivalent to four particle GHZ state and has a higher
entanglement persistency \cite{br01} and use the obtained cluster
state to implement some interesting quantum computation tasks. By
exploiting photon's intrinsic flying nature, we could also
envision that this experimental technique maybe applicable in
distributed quantum computation and ``quantum internet''
\cite{bmd04}. When combined with recent advance in neutral atoms
trapped in an optical lattice \cite{jbc99} and atom-photon
entanglement \cite{bmd04}, we could also dream of a
photon-assisted atomic one-way quantum computer that can
efficiently implement distributed quantum information processing.
We expect this work would stimulate further work towards feasible
quantum computation. In any event, the experimental results
present here may provide a first step towards that goal.

We thank Qiang Zhang and Hans Briegel for helpful discussions.
This work was supported by the National Natural Science Foundation
of China, Chinese Academy of Sciences and the National Fundamental
Research Program (underGrant No. 2001CB309303). This work was also
supported by the Alexander von Humboldt Foundation. 

\end{document}